\input{aipcheck}

\newcommand{\pt}{\ensuremath{p_{\rm t}}}
\newcommand{\pttrig}{\ensuremath{p_{\rm t, trig}}}
\newcommand{\ptassoc}{\ensuremath{p_{\rm t, assoc}}}
\newcommand{\kt}{\ensuremath{k_{\rm t}}}
\newcommand{\pp}{pp}
\newcommand{\PbPb}{Pb-Pb}

\documentclass[
  ,draft            
  ]
  {aipproc}

\layoutstyle{8x11single}

\begin{document}

\title{Reconstruction of Jet Properties in Pb-Pb Collisions with the ALICE-Experiment}


\author{C. Klein-B{\"o}sing for the ALICE Collaboration}{
  address={Institut f{\"u}r Kernphysik - M{\"u}nster, Germany}
 ,altaddress={ExtreMe Matter Institute, GSI - Darmstadt, Germany} 
}

\begin{abstract}

The study of the hot and dense medium created in heavy-ion collisions via 
the effect of parton energy loss is one of the major goals of jet and
high-\pt\ measurements in these reactions. In particular the
modifications of the longitudinal and
transverse jet structure. Here the subtraction of the underlying heavy-ion event, 
as well as a precise knowledge of the background induced fluctuations
of the reconstructed jet energy is required.  We extract
the modification of the near- and away-side jet yield in heavy-ion
collisions at the LHC and present a detailed assessment of 
background fluctuations for jet reconstruction.

\end{abstract}

\maketitle


\section{Introduction}

Hard scattered partons in heavy-ion collisions provide a well defined
probe for the complete evolution of the medium created in heavy-ion
collisions since the scattering occurs well before the medium is
formed ($t \approx 1/Q \ll 1$fm/$c$). Thus, the modification of the
final state fragmentation process of the parton compared to the
fragmentation into the QCD vacuum in \pp\ allows to map out the
properties of the QCD medium
\cite{Salgado:2002ws,Alessandro:2006yt}. Here, the most direct access
to the original parton properties is given by the full reconstruction
of jets. Already the first measurements at LHC revealed a striking
energy imbalance between back-to-back dijets \cite{Aad:2010bu},
pointing to a significant dissipation of jet energy in the medium. A
quantitative understanding of these results requires a precise
understanding of background-induced fluctuations of the measured jet
energy, which can distort the energy balance even in the absence of
any other medium effects \cite{Cacciari:2011tm}. It is also important
to corroborate full jet recoqnstruction measurements, with jet
properties extracted from single particle \cite{Aamodt:2010jd} and
correlation measurements, to take advantage of a direct fragmentation
bias and a different susceptibility to the underlying event
background. In case of the correlation measurement, this provides
access to jet properties in a momentum range. where full jet
reconstruction is dominated by background fluctuations.

\section{Hard Probes with the ALICE-Experiment}

The presented data were collected by the ALICE experiment
\cite{Aamodt:2008zz} during the first heavy-ion run of the LHC in the
fall of 2010 with lead ions colliding at an energy of $\sqrt{s_{\rm NN}} =                                                            
2.76$~TeV. The analyses are based on tracks of charged particles
reconstructed in the Time-Projection-Chamber (TPC) together with
vertexing information from the Inner Tracking System (ITS). 
This ensures maximum azimuthal angle ($\phi$) uniformity of
reconstructed tracks with transverse momenta down to \pt\  = 150 MeV/$c$.

\subsection{Triggered Particle Correlations}

\begin{figure}
\includegraphics[width=0.9\textwidth]{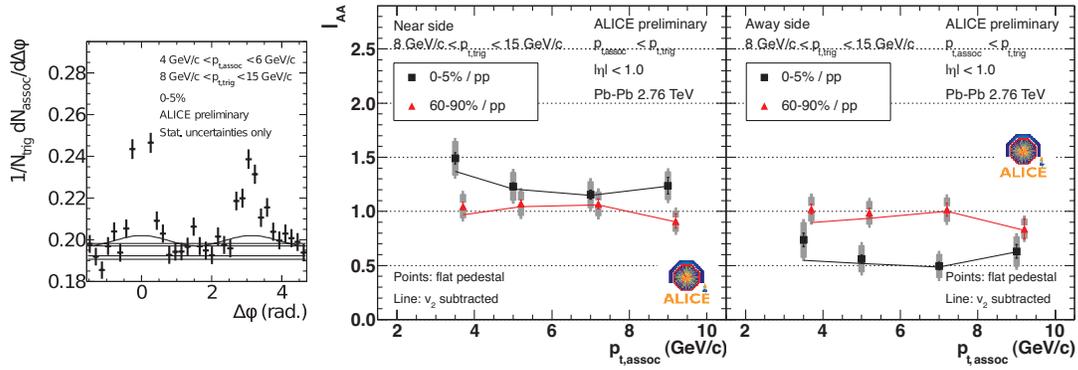}
  \caption{Left: Per-trigger yield (normalized by ($\Delta\eta = 1.6$) in an
    example bin (zoom, the near-side peak is off scale). The
    determined	pedestal values (horizontal lines) and the $v_2$
    component ($\cos 2\Delta\phi$) are shown. Right: $I_{AA}$ for near-
    and away-side in central and peripheral collisions: the data points are calculated with a flat pedestal; the line is based on $v_2$ subtracted yields.\label{fig:corr}}
\end{figure}

In this approach a trigger particle with $\pt > \pttrig$ is used to
define the direction for the measurement of azimuth difference
($\Delta\phi$) correlations of associated particles with $\ptassoc <
\pt < \pttrig$.  The requirement of a trigger particle prefers the
selection of unquenched jets close to the surface at the near-side,
while the away-side jets have to traverse a longer path through the
medium. At lower \pttrig\ where particle production is not dominated
by jet fragmentation, collective effects \cite{Aamodt:2010pa}, have a
strong influence on the $\Delta\phi$ distributions and a
Fourier-decomposition can be used to quantify the role of long range
correlations (see e.g. \cite{GrosseOetringhaus:2011kv}).  Here, the
trigger particle \pt\ has been chosen such that collective effects on
the $\Delta\phi$ distribution are small and jet-like correlations
dominate ($\pttrig > 8$~GeV/$c$, $\ptassoc > 3$~GeV/$c$). The
distribution of associated particles with respect to the $\phi$
direction of the trigger particle is normalized to the number of
triggers as illustrated in Figure~\ref{fig:corr}, where a clear
back-to-back structure at a distance $\Delta\phi = \pi$ is seen. As
illustrated in Figure~\ref{fig:corr}, the background (i.e. non-jet)
contribution to the raw trigger yield is subtracted using different
assumptions on the magnitude of a constant pedestal or a constant
pedestal plus the expected contribution from elliptic flow, which
leads to an additional $\cos(2\Delta\phi)$ variation. After background
subtraction the per-trigger-yield $Y$ is obtained by integrating the
near- and away-side peak, respectively. To quantify the modification
of the jet fragmentation process in different bins of $\ptassoc$, the
ratio $I_{AA} = Y_{\PbPb}/Y_{\pp}$ to the yield measured in \pp\ at
$\sqrt{s} = 2.76$~TeV is used. As seen in Figure~\ref{fig:corr}, for
peripheral collisions (60-90\%) neither the yield on the away-side,
nor on the near-side is modified compared to elementary
reactions. However, for central (0-5\%) a significant suppression of
the away-side yield is observed for all bins of \ptassoc\, as expected
from parton energy loss.  Furthermore, the yield at the near-side is
enhanced, which can be interpreted as a consequence of jet quenching
as well. One effect is that near-side gluons contribute less to
the fragmentation yield, since they are more effectively quenched in
the medium, due to their larger color charge. Quarks fragment harder
and thus can give rise to the enhanced yield \cite{Renk:2011wp}.

\subsection{Jet Background Characterization}

The separation of jet properties from the global event characteristics, in
the case of triggered correlation achieved via a large \pttrig\,
becomes more challenging in the case of full jet reconstruction. Here,
the goal is to recover the full parton momentum by the use of various
algorithms which cluster the final state particles to a jet (here the
sequential recombination algorithms \kt\ and {anti-\kt} from the
FastJet package are employed \cite{Cacciari:2005hq}).
From the jet momentum and the particles within the jet in principle
the fragmentation function of the parton and its medium modification
can be reconstructed. However, in the context of heavy-ion collisions
one has to take into account the significant contribution of the
underlying event to the reconstructed jet momentum, which can be
reduced e.g. by a \pt\ or energy threshold of the input particles to
the jet finders and by a reduced distance parameter $R$. All of these
introduce additional biases on the fragmentation pattern of the
reconstructed jets and the remaining background has
to be subtracted from the reconstructed jet \pt. For the present
analysis the background
subtraction scheme proposed in \cite{Cacciari:2007fd} is employed,
which uses an event-by-event median of the background momentum per
unit area $\rho$, which accounts for the expected contribution in a
jet with area $A$. In the following the fluctuations induced
by correlated and uncorrelated variations of the background density
within one event are investigated. This measurement is of particular
importance for the correction of the inclusive jet spectrum and for
the quantitative interpretation of jet momentum imbalance. The current
study is limited to the charged component of the jet, and will be
extended in the next \PbPb\ run by ALICE with the addition of the
EMCAL to the combined charged and neutral particle information.
\begin{figure}
\includegraphics[width=0.89\textwidth]{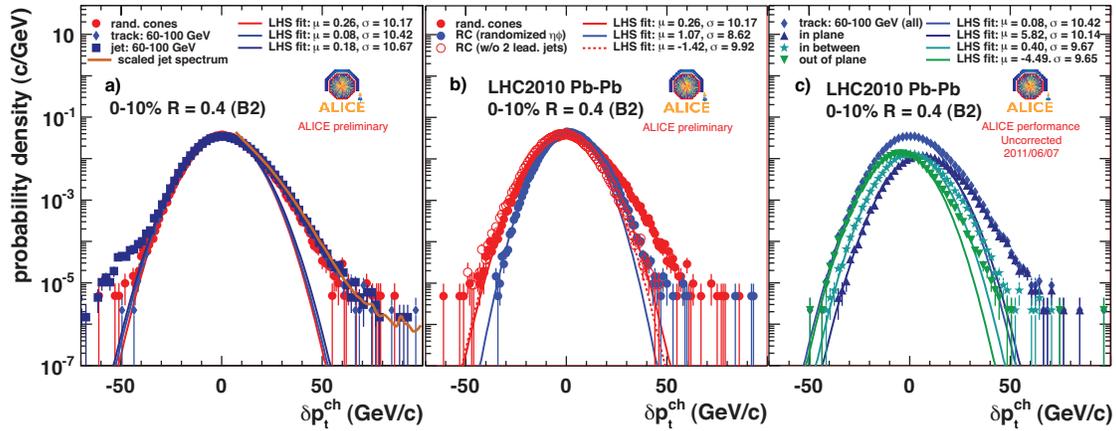}
  \caption{$\delta\pt$ distributions for reconstructed charged
    particles with $\pt > 150$~MeV/$c$ and
    different probes types. The left-hand-side of the distributions is fit to a
    Gaussian that is plotted over the full range to indicate the
    deviation from the Gaussian shape.\label{fig:fluc}}
\end{figure}

The fluctuations are quantified by embedding well defined probes into
the measured heavy-ion events and calculating the residuals after background correction
$\delta\pt = \pt^{\rm rec} - \rho A - \pt^{\rm probe}$.
The different probes involve: (i) random cones (RC) of fixed area placed
into the acceptance, (ii) high-\pt\ particles acting as effective
seeds for jet finding at this position (iii) \pp\ jets from a full
PYTHIA+GEANT simulation. The different cases are shown in
Figure~\ref{fig:fluc}a) for the 0-10\% most central events. They are
centered at zero, showing the validity of our background subtraction
scheme and show in general a good agreement amongst each other. The
deviations from a purely Gaussian shape is clearly seen, the width of
the Gaussian is in all cases larger than 10 GeV/$c$. One source of
the deviation is the presence of jets in the event: this can enhance  the
local \pt\ density in the rare case of a hard scattering significantly. In Figure~\ref{fig:corr}b), random
cones are embedded into real events for two additional cases: with a distance of $D > 1$ in
the $\eta\phi$-plane to the two leading jets and after randomizing the
momentum vectors of the event in the acceptance. In both cases the
tail on the right-hand-side attributed to jets is much reduced. In the
latter case also the width on the left side is reduced, pointing to
correlated fluctuation not originating from jets. These can be
explained by the presence of collective flow, which modulates the
particle density and the average momentum in a heavy-ion event
depending on the direction to the reaction plane (given by the beam
direction and the impact parameter of the colliding nuclei).
Figure~\ref{fig:corr}c) illustrated this with the shift of the
$\delta\pt$ distribution resulting from an over and under-subtraction
of the background, when embedding into different direction with
respect to the reaction plane.

\section{Summary}

We presented  the modification of the near- and
away-side jet yield measured via triggered hadron correlations in
heavy-ion
collisions at the LHC and also showed an assessment of the background
fluctuations for jet reconstruction in heavy-ion collisions, which
exceed 10 GeV/$c$ for charged particles in central \PbPb\ reactions.


\begin{theacknowledgments}
Copyright CERN for the benefit of the ALICE Collaboration.
\end{theacknowledgments}



\bibliographystyle{aipproc}   

\bibliography{5K5_Klein-Boesing}

\end{document}